# Universal effective coupling constants for the generalized Heisenberg model

A. I. Sokolov

*St. Petersburg State Electrotechnical University, 197376 St. Petersburg, Russia*

The aim of this study is to find universal critical values of the effective dimensionless coupling constant $g_6$ and refined universal values $g_4$ for Heisenberg ferromagnets with $n$-component order parameters. These constants appear in the equation of state and determine the nonlinear susceptibilities $\chi_4$ and $\chi_6$ in the critical region. Calculations are made of the first three terms of the expansion of $g_6$ in powers of $g_4$ in the limits of $O(n)$ symmetry three-dimensional $\lambda\varphi^4$ theory, the resultant series is resummed by the Padé–Borel method, and then by substituting the fixed point coordinates $g_4^*$ in the resultant expression, numerical values of $g_6^*$ are obtained for different $n$. These numbers $g_4^*$ for $n>3$ were determined from a six-loop expansion for the $\beta$-function resummed using the Padé–Borel–Leroy technique. An analysis of the accuracy of these $g_6^*$ values showed that they may differ from the true values by no more than 1.6%. These values of $g_6^*$ were compared with those obtained by the $1/n$ expansion method which allowed the level of accuracy of this method to be assessed.

The generalized Heisenberg model, which consists of a lattice of $n$-dimensional spins, each interacting only with its nearest neighbors, occupies a central position in the theory of phase transitions. It describes critical phenomena in a wide range of objects including easy-axis, easy-plane, and Heisenberg ferromagnetics ($n=1,2,3$), simple liquids and binary mixtures ($n=1$), superconductors (except for heavy-Fermion and, obviously, high-temperature superconductors), and superfluid helium-4 ($n=2$). This model describes the limiting regimes of critical behavior of two superfluid Fermi liquids with triplet pairing: helium-3 ($n=18$) (Refs. 1 and 2) and neutron-star matter ($n=10$) (Refs. 3 and 4), and also a quark–gluon plasma in various models of quantum chromodynamics ($n=4$) (Refs. 5 and 6).

It is known that the generalized Heisenberg model is thermodynamically equivalent in the critical region to the classical $O(n)$-symmetric three-dimensional Euclidean field theory with $\lambda\varphi^4$ interaction. This means that quantum-field theory and, in particular, the renormalization group method, which has proved exceptionally effective in analyses of the qualitative features of critical behavior and also in calculations of critical exponents,[7–9] can be used to study its critical properties. However, the critical exponents are not the only fundamental parameters characterizing the thermodynamics of a system in the strong fluctuation range. Equally important are the effective dimensional coupling constants $g_{2k}$ which appear in the equation of state and determine the nonlinear susceptibilities of different orders.

In recent years, the problem of finding the universal critical values $g_6$, $g_8$, and other higher-order coupling constants has attracted particular attention.[10–21] A whole range of available methods have been applied to solve this problem, ranging from the purely analytical[14,16–18] to the Monte Carlo method.[12,21] However, the theoretical activity has been almost exclusively confined to the case $n=1$, i.e., the Ising model. There is only one study where the universal values of $g_6$ were determined for $n>1$ (Ref. 13) but the accuracy achieved (12–24%) can hardly be considered to be satisfactory.

It was observed quite recently that extremely accurate numerical estimates for $g_6^*$ may be obtained using the theoretical-field renormalization group method[16–18] in fairly low orders of perturbation theory. In fact, calculations of $g_6^*$ for $n=1$ in three-, four-, and five-loop approximations made by resumming the renormalization group expansions for the three-dimensional model yielded 1.622 (Ref. 16), 1.596 (Ref. 17), and 1.604 (Ref. 18), respectively. The last of these values, being the most accurate, only differs from its three-loop analog by 1.1%. However, as $n$ increases, the suitably normalized coefficients of the renormalization group expansions decrease (see Ref. 9, for instance) which leads to an improvement in the approximating properties of these series. Thus, for $n>1$ the three-loop renormalization group expansions for $g_6$ should give numerical estimates whose level of accuracy is better than 1–2% in any case. In this situation, it is natural to use the theoretical-field renormalization group technique in three-dimensional space to calculate the universal critical values of $g_6$ for arbitrary dimensionality of the order parameter, and this is the problem addressed in the present paper.

This paper is organized as follows. Section 1 contains general information needed to formulate the problem and to derive the renormalization group expansion for the effective coupling constant $g_6$. In Sec. 2, a six-loop expansion of the $\beta$ function is used as the basis to calculate the coordinate of the nontrivial fixed point $g_4^*$ for $n>3$. In this case, a Borel–Leroy transformation and several different types of Padé approximants are used to resum the renormalization group series, which can give more accurate numerical values of $g_4^*$ than those obtained previously.[9] In Sec. 3, universal critical asymptotic forms of $g_6$ are obtained for different $n$, the results are compared with those obtained by the $1/n$ expansion, and an analysis is made.

## 1. RENORMALIZATION GROUP EXPANSION FOR THE EFFECTIVE COUPLING CONSTANT $g_6$

The Hamiltonian of this model has the form

$$H = \int d^3x \left[ \frac{1}{2}(m_0^2 \varphi_\alpha^2 + (\nabla \varphi_\alpha)^2) + \lambda(\varphi_\alpha^2)^2 \right], \quad (1)$$

where $\varphi_\alpha$ is the real $n$-component vector field, the "bare mass" squared $m_0^2$ is proportional to $T - T_c^{(0)}$, and $T_c^{(0)}$ is the phase transition temperature neglecting fluctuations. Allowance for fluctuations leads to renormalization of the mass $m_0^2 \to m^2$, the field $\varphi_\alpha \to \varphi_{\alpha R}$, and the coupling constant $\lambda \to m g_4$, and also results in the appearance of higher-order terms in the expansion of the free energy in powers of the magnetization $M$

$$F(M,m) = F(0,m) + \sum_{k=1}^{\infty} \Gamma_{2k} M^{2k}. \quad (2)$$

The expansion coefficients $\Gamma_{2k}$ comprise complete vertices with $2k$ external (truncated) lines which are linked by simple relations with $2n$-point 1-irreducible correlation functions $G_{2k}(q_1, q_2, \ldots, q_{2n-1})$ at zero momenta. In the critical region each vertex has its scale dimensionality

$$\Gamma_{2k} = g_{2k} m^{3-k(1+\eta)}, \quad (3)$$

where $\eta$ is the Fisher index and $g_{2k}$ are various constants. The first of these $g_2$ is arbitrary in the sense that its value can be fixed by selecting the units of measurement of the renormalized mass $m$. We assume as usual that $g_2 = 1/2$ and then $m$ is the same as the reciprocal correlation radius and the linear susceptibility $\chi_2$ in the disordered phase will be equal to $m^{\eta-2}$. The second constant $g_4$ is a key parameter of the theory, in whose terms the critical exponents, the critical amplitude ratios, and other universal characteristics of the system are expressed. The asymptotic value $g_4$ (the coordinate of the fixed point $g_4^*$) is a nontrivial root of the $\beta$ function appearing in the renormalized group equation. For the model (1) this function is known in the highest-reported six-loop approximation,[7–9] so that $g_4^*$ can be found for any $n$ with extremely low error.

The higher effective coupling constants $g_6$, $g_8$, and others also have certain universal values in the limit $T \to T_c$ which jointly determine the form of $F(M,n)$ and the equations of state in the strong fluctuation region. In reality, however, the Taylor expansion of the scaling function normally used to write the equation of state contains not these constants but the ratio $g_{2k}/g_4^{k-1}$, which can easily be seen by replacing the magnetization $M$ in Eq. (2) by the dimensional variable $z = M\sqrt{g_4/m^{(1+\eta)}}$:

$$F(z,m) - F(0,m) = \frac{m^3}{g_4}\left(\frac{z^2}{2} + z^4 + \frac{g_6}{g_4^2}z^6 + \frac{g_8}{g_4^3}z^8 + \ldots\right). \quad (4)$$

Nonlinear susceptibilities of different orders can then be expressed directly in terms of $g_{2k}$. For $\chi_4$ and $\chi_6$, for example we can easily derive the following formulas:

$$\chi_4 = \left.\frac{\partial^3 M}{\partial H^3}\right|_{H=0} = -24\chi_2^2 m^{-3} g_4,$$

$$\chi_6 = \left.\frac{\partial^5 M}{\partial H^5}\right|_{H=0} = 720\chi_2^3 m^{-6}(8g_4^2 - g_6), \quad (5)$$

whose generalization yields the relations

$$g_4 = -\frac{m^3 \chi_4}{24\chi_2^2}, \quad g_6 = \frac{m^6(10\chi_4^2 - \chi_6\chi_2)}{720\chi_2^4}, \quad (6)$$

which are frequently used to determine the dimensionless effective coupling constants using the results of lattice calculations.[13,15,20,22,23]

We shall now find the renormalization group expansion for $g_6$. We shall start from normal perturbation theory which gives a diagram series for $\Gamma_6$. Since in three-dimensional space only $\lambda \varphi^4$ type interaction is important in the renormalization group sense (see Ref. 24), only four-point functions will appear as seed vertices in the diagrams of this series. Given the expansion of $\Gamma_6$ in powers of $\lambda$, we can then renormalize it, expressing $\lambda$ in terms of $g_4$ using the familiar relation

$$\lambda = m Z_4 Z^{-2} g_4, \quad (7)$$

where $Z_4$ and $Z$ are the renormalization constants of the interaction $\lambda$ and the field $\varphi_\alpha$: $\varphi_\alpha = \sqrt{Z}\varphi_{\alpha R}$. After replacing $\Gamma_g$ by $g_6$, this procedure gives the required result.

In the three-loop approximation, the vertex $\Gamma_6$ is reduced to the sum of the contributions of twenty Feynman diagrams which are shown in Fig. 1. The integrals corresponding to these diagrams can easily be calculated but when determining the tensor factors, it should be borne in mind that all the vertices in the model (1) are symmetric tensors of the corresponding ranks. The tensor structure of the objects of interest to us is determined in particular, by the following formulas:

$$\Gamma_{\alpha\beta\gamma\delta} = \frac{1}{3}(\delta_{\alpha\beta}\delta_{\gamma\delta} + \delta_{\alpha\gamma}\delta_{\beta\delta} + \delta_{\alpha\delta}\delta_{\beta\gamma})\Gamma_4, \quad (8)$$

$$\Gamma_{\alpha\beta\gamma\delta\mu\nu} = \frac{1}{15}(\delta_{\alpha\beta}\delta_{\gamma\delta}\delta_{\mu\nu} + 14 \text{ transpositions})\Gamma_6. \quad (9)$$

Thus, calculations of the diagrams (Fig. 1) give

$$g_6 = \frac{9}{\pi}\left(\frac{\lambda Z^2}{m}\right)^3 \left[\frac{n+26}{27} - \frac{9n^2 + 340n + 2324}{162\pi}\left(\frac{\lambda Z^2}{m}\right)\right.$$

$$+ (0.00562895 n^3 + 0.28932673 n^2 + 4.04042412 n$$

$$\left. + 16.20428685)\left(\frac{\lambda Z^2}{m}\right)^2\right]. \quad (10)$$

The expansion of the renormalization constant $Z_4$ for the model (1) is now known in the sixth order in $g_4$ (Ref. 25) but we only require the first three orders of this term

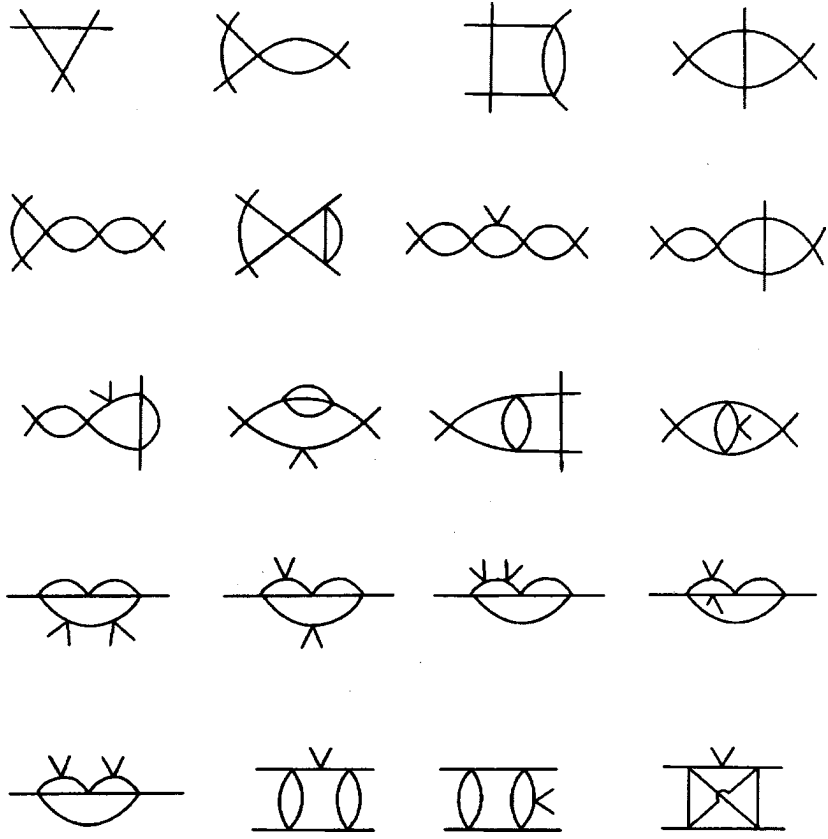

FIG. 1. One-, two-, and three-loop Feynman diagrams which contribute to the effective coupling constant $g_6$.

$$Z_4 = 1 + \frac{n+8}{2\pi} g_4 + \frac{3n^2 + 38n + 148}{12\pi^2} g_4^2. \quad (11)$$

Substituting Eq. (11) into Eq. (7), and then Eq. (7) into Eq. (10) finally gives

$$g_6 = \frac{9}{\pi} g_4^3 \left[ \frac{n+26}{27} - \frac{17n+226}{81\pi} g_4 + (0.00099916 n^2 + 0.14768927 n + 1.24127452) g_4^2 \right]. \quad (12)$$

This renormalization group expansion will be used to calculate the universal critical values of $g_6$.

## 2. FIXED POINT COORDINATES FOR $n > 3$

In order to find the asymptotic values of $g_6$ for different $n$, we need to know the fixed-point coordinates of the renormalization group equation with the highest possible accuracy. At the present time, this highest possible accuracy can be achieved by applying various resummation procedures to the six-loop expansion of the $\beta$-function of the model (1). Two decades ago, this method was used to determine the numerical values of $g_4^*$ for $n = 1,2,3$ (Refs. 7 and 8) and comparatively recently for $n > 3$ (Ref. 9). However, whereas the authors of Refs. 7 and 8 used complex, refined procedures for summation of divergent series based on the Borel–Leroy transformation and various methods of analytic continuation, especially using the conformal mapping technique,[8] the authors of Ref. 9 confined themselves to a simple Borel approximation and Padé approximants of only one type $[L-1/1]$. It can be seen that, in principle, this method gives reasonable results: for $n > 10$ the difference between the values of $g_4^*$ obtained by numerical estimates and their analogs obtained by more complex resummation procedures does not exceed 0.001 (0.1%). However, for smaller $n$ this difference is appreciable, causing us to search for refined values of $g_4^*$.

Thus, the expansion of the $\beta$ function of the model (1) in the six-loop approximation has the form[9]

$$\beta(g) = g - g^2 + \frac{1}{(n+8)^2}(6.07407408 n + 28.14814815) g^3 - \frac{1}{(n+8)^3}(1.34894276 n^2 + 54.94037698 n + 199.6404170) g^4 + \frac{1}{(n+8)^4}(-0.15564589 n^3 + 35.82020378 n^2 + 602.5212305 n + 1832.206732) g^5 - \frac{1}{(n+8)^5}(0.05123618 n^4 + 3.23787620 n^3 + 668.5543368 n^2 + 7819.564764 n$$

$$+20770.17697)g^6 + \frac{1}{(n+8)^6}(-0.02342417n^5$$
$$+1.07179839n^4 + 265.8357032n^3$$
$$+12669.22119n^2 + 114181.4357n$$
$$+271300.0372)g^7. \quad (13)$$

Here, as in previous studies,[7–9] the role of the argument is not played by the effective coupling constant $g_4$ but by the dimensionless invariant charge proportional to it

$$g = \frac{n+8}{2\pi}g_4, \quad (14)$$

which unlike $g_4$, does not tend to zero for $n \to \infty$ but reaches a final value of unity. Series of the type (13) are known to be asymptotic but they can easily be reduced to convergent series by means of a Borel–Leroy transformation

$$f(x) = \sum_{i=0}^{\infty} c_i x^i = \int_0^{\infty} e^{-t} t^b F(xt) dt,$$

$$F(y) = \sum_{i=0}^{\infty} \frac{c_i}{(i+b)!} y^i. \quad (15)$$

In order to calculate the integral in Eq. (15), we need to analytically continue the Borel transform $F(y)$ of the unknown function beyond the circle of convergence. To do this, we can use the Padé approximants [L/M] which are the ratios of the polynomials $P_L(y)$ and $Q_M(y)$ of orders $L$ and $M$ whose coefficients are determined uniquely if $L+M+1$ is equal to the number of known terms in the series and $Q_M(0)=1$. It was established that the best approximating properties are exhibited by diagonal Padé approximants for which $L=M$ or close to them (c.f., Ref. 26). However, as the degree of the denominator $M$ increases, the number of its roots increases, i.e., the number of poles of the approximant in the complex plane. If at least some of these poles are located near the real semiaxis $y>0$ or, which is even worse, lie on it, the corresponding approximant becomes unsuitable for summation of the series. In practice, this imposes a fairly stringent upper constraint on the degree of the denominator and narrows the choice of acceptable approximants. However, the presence of a free parameter $b$ in the Borel–Leroy transformation allows the resummation procedure to be optimized to achieve the fastest possible convergence of the iteration scheme.

With these observations in mind, we selected the following strategy to calculate $g^*$ ($g_4^*$). For each $n$ the nontrivial root of the equation $\beta(g)=0$ was found in the two leading approximations — five-loop and six-loop — and the Borel transforms of the $\beta$ function were continued analytically by means of three types of Padé approximants: [3/3], [4/2], and [3/2]. The values of the parameter $b$ were varied widely (usually between 0 and 30) and were selected so that the numerical results given by the five- and six-loop expansions were the same or very close for all these types of approximants, i.e., the fastest convergence of the iteration procedure was ensured. In those cases where the diagonal approximant [3/3] had poles for positive or small negative $y$ for all reasonable values of $b$, the value of $g^*$ was determined using the other two less symmetric approximants [4/2] and [3/2] in their range of analyticity. When these approximants also became unsuitable with increasing $n$ because of the appearance of ''dangerous'' poles (this occurred between $n=28$ and $n=32$), the approximants used to obtain estimates for $g_4^*$ were switched to [5/1] and [4/1], which could still be used up to $n=40$. The fixed-point coordinates thus obtained and also the values of the critical exponent $\omega = -d\beta(g^*)/dg$, which determines the temperature dependences of the corrections to the scaling, are given in Table I (columns 1 and 2). Also given for comparison are the values of $g^*$ obtained earlier[9] by the Padé–Borel method using the [5/1] approximant (column 3) and also by using more refined methods of resummation[7,8] (columns 4 and 5).

TABLE I. Fixed-point coordinates $g^*$, critical exponent $\omega$, and universal values of the coupling constant $g_6$ for $1 \leq n \leq 40$.

| n | $g^*$ | $\omega$ | $g^*$ (Ref. 9) | $g^*$ (Ref. 7) | $g^*$ (Ref. 8) | $g_6^*$ | $g_6^*$ (Ref. 13) | $g_6^*(1/n)$ |
|---|---|---|---|---|---|---|---|---|
|  | 1 | 2 | 3 | 4 | 5 | 6 | 7 | 8 |
| 1 | 1.419 | 0.781 | 1.401 | 1.416 | 1.414 | 1.622 | 1.925±0.242 | |
| 2 | 1.4075 | 0.780 | 1.394 | 1.406 | 1.405 | 1.236 | 1.268±0.246 | |
| 3 | 1.392 | 0.780 | 1.383 | 1.392 | 1.391 | 0.956 | 0.933±0.197 | 2.9243 |
| 4 | 1.3745 | 0.783 | 1.369 | | | 0.751 | 0.621±0.146 | 1.6449 |
| 5 | 1.3565 | 0.788 | 1.353 | | | 0.599 | | 1.0528 |
| 6 | 1.3385 | 0.793 | 1.336 | | | 0.485 | | 0.7311 |
| 7 | 1.321 | 0.800 | 1.319 | | | 0.398 | | 0.5371 |
| 8 | 1.3045 | 0.808 | 1.303 | | | 0.331 | | 0.4112 |
| 9 | 1.289 | 0.815 | 1.288 | | | 0.278 | | 0.3249 |
| 10 | 1.2745 | 0.822 | 1.274 | | | 0.236 | | 0.2632 |
| 12 | 1.2487 | 0.836 | 1.248 | | | 0.175 | | 0.1828 |
| 14 | 1.2266 | 0.849 | 1.226 | | | 0.134 | | 0.1343 |
| 16 | 1.2077 | 0.861 | 1.207 | | | 0.105 | | 0.1028 |
| 18 | 1.1914 | 0.871 | 1.191 | | | 0.0847 | | 0.0812 |
| 20 | 1.1773 | 0.880 | 1.1768 | | | 0.0694 | | 0.0658 |
| 24 | 1.1542 | 0.896 | 1.1538 | | | 0.0488 | | 0.0457 |
| 28 | 1.1361 | 0.909 | 1.1359 | | | 0.0361 | | 0.0336 |
| 32 | 1.1218 | 0.919 | 1.1216 | | | 0.0276 | | 0.0257 |
| 36 | 1.1099 | 0.927 | 1.1099 | | | 0.0218 | | 0.0203 |
| 40 | 1.1003 | 0.934 | 1.1003 | | | 0.0176 | | 0.0164 |

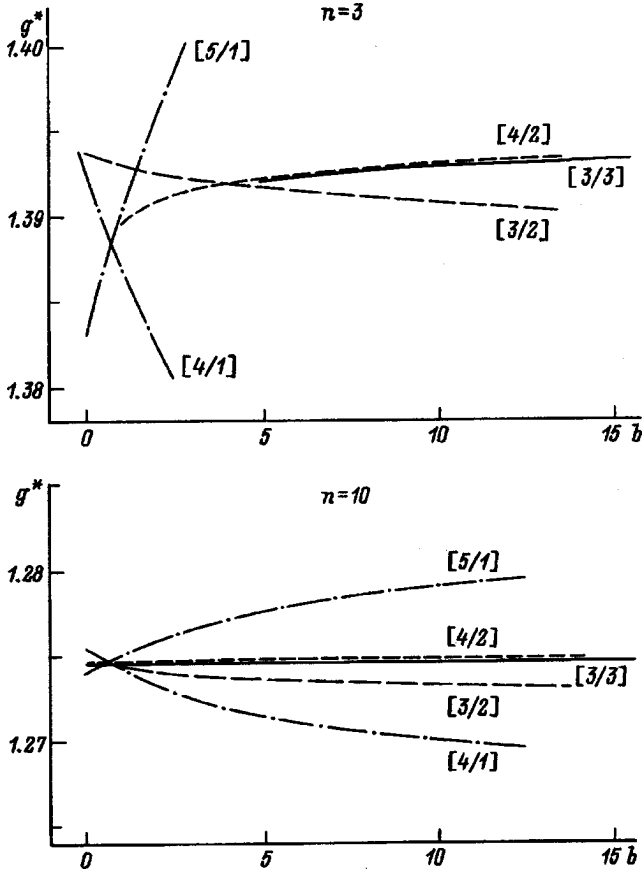

FIG. 2. Nontrivial fixed-point coordinates for $n=3$ and 10 calculated by the Padé–Borel–Leroy method using five different types of Padé approximants, as a function of the parameter $b$.

A comparison between the numbers in columns 1, 4, and 5 can be used to test this algorithm. For instance, for $n=3$ the difference between this estimate of $g^*$ and the more accurate ones does not exceed 0.001. Another argument in support of the efficiency of this technique may be that the numerical estimates given by the main ''working'' approximants [3/3] and [4/2] depended very weakly on the parameter $b$. This is clearly illustrated in Fig. 2 which gives the dependences $g^*(b)$ for $n=3$ and 10 obtained by using five different Padé approximants. It can be seen that for $n=3$ the values of $g^*$ calculated using the [3/3] and [4/2] approximants increased only by 0.0015 when $b$ increased from 5 to 15 and for $n=10$ their increase was less than 0.0003 in the range $0<b<15$. Since the fixed-point coordinates given by the [3/3] and [4/2] approximants are almost the same (Fig. 2), the errors in the determination of $g^*$ in any case should not exceed the ranges of variation of these values given above.

In the next section the refined values of the nontrivial fixed-point coordinates will be used to find the critical asymptotic forms of $g_6$ for various $n$.

## 3. UNIVERSAL VALUES OF THE EFFECTIVE COUPLING CONSTANT $g_6$

We shall now determine the universal critical values of $g_6$. Using Eqs. (12) and (14), we express the coupling constant $g_6$ in terms of the charge $g$

$$g_6 = \frac{8\pi^2}{3} \frac{n+26}{(n+8)^3} g^3 \left[ 1 - \frac{2(17n+226)}{3(n+8)(n+26)} g + \frac{(1.065025 n^2 + 157.42454 n + 1323.09596)}{(n+8)^2(n+26)} g^2 \right]. \quad (16)$$

As $n$ increases, the coefficients of $g$ and $g^2$ in formula (16) decrease and in addition, the expansion parameter $g^*$ also decreases, as can be seen from Table I. Consequently, the approximating properties of this series should improve with increasing $n$. For $n=1$, summation of the expansion (16) by the Padé–Borel method using the diagonal [1/1] approximant yielded the estimate $g_6^* = 1.622$ (Ref. 16) which only differs by 0.018 from the result $g_6^* = 1.604$ (Ref. 18) obtained using the five-loop approximation. As has been noted, the similarity between these numbers is not coincidental and reflects the rapid convergence of the iteration process. In this situation, it is quite natural to use the tried summation technique to calculate $g_6^*$ for arbitrary $n$. Thus, by constructing the series for the Borel transform of the function $g_6(g)$ in accordance with Eq. (15), continuing its sum analytically using a Padé [1/1] approximant, and substituting into the resulting expression $g = g^*$, we can easily obtain the universal critical values of $g_6$ which are given in column 6 of Table I. With this in mind, we can confirm that these values differ by no more than 1.1% from the $g_6^*$ values given by the five-loop renormalization group expansion. However, the accurate values of $g_6^*$ should lie between the four- and five-loop estimates since the series for $g_6$ is alternating. Since for $n=1$ the four-loop approximation gives $g_6^* = 1.596$ (Ref. 17), the differences between the numbers in column 6 and the true critical values of $g_6$ cannot exceed 1.6%.

It is interesting to compare these values with those obtained by Reisz for $n=1,2,3,4$ using lattice expansions[13] which are given in column 7. It can be seen that although the results of Ref. 13 differ appreciably from those obtained here, no direct contradiction exists (except for the case $n=1$). A second interesting exercise is to compare the estimates obtained for $g_6^*$ with those obtained by the $1/n$ expansion method. Summing all contributions of order $1/n^2$, i.e., replacing the ''bare'' vertex in the first graph in Fig. 1 by sums of ladder diagrams, we obtain

$$g_6^* = \frac{8\pi^2}{3n^2} + O\left(\frac{1}{n^3}\right). \quad (17)$$

The numerical values of $g_6^*$ given by this formula are listed in the last column of Table I. On comparing the values given in columns 6 and 8, it is easy to see that the $1/n$ expansion being applied to find $g_6^*$, gives considerably inferior results compared with calculations of the fixed-point coordinate and the critical exponents. Whereas in this last case, a 1% level of accuracy is still achieved for $n=28$ (Ref. 9), in calculations of $g_6^*$ even for $n=40$ the accuracy of the estimates obtained using the $1/n$ expansion is 6% worse. The almost exact agreement between the values of $g_6^*$ from columns 6 and 8 for $n=14$ does not alter this conclusion since for this value of $n$ the curves $g_6^*(n)$ given by the resummed renormalization group expansion (16) and Eq. (17) simply intersect.

To conclude, another observation should be made regarding the accuracy of these results. An error in the determination of $g_6^*$ not exceeding 1.6% can undoubtedly be considered fairly small. However, experimental technology has recently developed so rapidly that it is now possible to measure critical exponents to the fourth decimal place.[27,28] It is quite feasible that a similar level of accuracy will soon be achieved in experimental determinations of the equations of state of systems described by the model (1). Thus, it is highly desirable to calculate $g_6^*$ in the next order of renormalized perturbation theory. Specifically, allowance for the four-loop contribution to $g_6$ would reduce the error in the calculation of its universal critical value at least threefold. At the same time, calculations of $g_6$ in the five-loop and higher renormalization-group approximations are obviously pointless at the present time. This is because even for $n=1$ the difference between the four- and five-loop estimates for $g_6^*$ is so small that it is completely obscured by the variation of the fixed-point coordinate $g^*$ within its determination error. Thus, allowance for the five-loop contribution to $g_6$ can only realistically improve the accuracy with which $g_6^*$ is calculated provided that its universal charge value $g$ is determined at the very least from a seven-loop expansion for the $\beta$ function. Such an expansion is as yet unknown.

The author would like to thank student S. S. Kashtanov for making some of the control calculations.

This work was supported by the Foundation for Intellectual Collaboration (St. Petersburg) as part of the Russian Scientific-Technical Program ''Fullerenes and Atomic Clusters'' and the Ministry of General and Professional Education of the Russian Federation (Grant No. 97-14.2-16).